\documentclass[preprint,12pt]{aastex} 

\author{Eric C. Bellm\altaffilmark{1,2}}
\title{Volumetric Survey Speed: \\ 
A Figure of Merit for Transient Surveys}

\altaffiltext{1}{Cahill Center for Astronomy and Astrophysics, California
Institute of Technology, Pasadena, CA 91125}
\altaffiltext{2}{ebellm@caltech.edu}

\usepackage{amsmath}
\usepackage{amssymb}
\usepackage[svgnames,pdftex,hyperref]{xcolor}
\usepackage[pdftex]{hyperref}
\hypersetup{colorlinks=True,linkcolor=Black,citecolor=Black,
urlcolor=Black,pagebackref,pdftitle={The Zwicky Transient Facility}
,pdfauthor={Eric Bellm}}

\usepackage{algorithm}
\usepackage{algpascal}

\usepackage{savesym}
\savesymbol{tablenum}
\usepackage{siunitx}
\restoresymbol{SIX}{tablenum}

\shorttitle{Survey Speed}
\shortauthors{Bellm}

\begin{abstract}
Time-domain surveys can exchange sky coverage for revisit frequency,
complicating the comparison of their relative capabilities.
By using different revisit intervals, 
a specific camera may execute surveys
optimized for discovery of different classes of transient objects.
We propose a new figure of merit, the instantaneous 
volumetric survey speed, for evaluating transient surveys.  
This metric defines the trade between cadence interval and snapshot
survey volume and so provides a natural means of comparing survey
capability.
The related metric of areal survey speed imposes a constraint on the range
of possible revisit times: we show that many modern
time-domain surveys are limited by the amount of fresh sky available each
night.
We introduce the concept of ``spectroscopic
accessibility'' and discuss its importance for transient science goals
requiring followup observing.
We present an extension of the control time algorithm for cases where
multiple consecutive detections are required.
Finally, we explore how survey speed and choice of cadence interval 
determine the detection rate of transients in the peak
absolute magnitude--decay timescale phase space. 
\end{abstract}

\keywords{telescopes, methods: observational}

\begin{document}
\maketitle

\section{Introduction}

Figures of merit provide a means of comparing and optimizing astronomical
instruments and surveys.  Good figures of merit encapsulate key capabilities,
contain relatively few assumptions, and compute easily with readily accessible
information.

For imaging surveys, the standard figure of merit is \'{e}tendue:
the product of a camera's field of view $\Omega_{\rm fov}$ and
the telescope's collecting area $A$.  When comparing sites of
varying image quality, it is common to normalize the \'{e}tendue by the square
of the FWHM of the point spread function
\citep[e.g.,][]{Terebizh:11:TelescopeDesigns}.  \'{E}tendue is then
proportional to the time needed to survey a large area of sky to a
specified depth.  Notably, it does not matter whether the depth is reached
by a deep single exposure (as from an instrument with large collecting area
but small field of view) or many shallower exposures (as from wide field
cameras on smaller telescopes).

For time-domain astrophysics, however, the depth and temporal sequence
of the exposures (their cadence)
are critical to determine what phenomena are detectable and
amenable to followup observations.  A single deep exposure is clearly not
equivalent to many shallower exposures when searching for supernovae, for
instance.  A new figure of merit is thus needed to compare the capabilities
of surveys in detecting transient astrophysical events.

One challenge in formulating such a figure of merit is that a given
instrument may execute surveys using a wide range of revisit times.  
For a fixed amount of total observing time, changing the time between
revisits to each field also changes the 
sky area it is possible to cover in each cadence interval.  This choice
then determines the discovery rate that is 
possible for various types of transient events.
For ease of comparison, however, we would like a figure of merit that is
independent of the survey strategy implemented.  We seek a metric derived
from the fundamentals of the camera, telescope, and site that illuminates
these trades.

\defcitealias{Tonry:11:ATLAS}{T11} 
\citet[\citetalias{Tonry:11:ATLAS}]{Tonry:11:ATLAS}
discussed these issues and 
proposed a capability metric derived from
the information theory of signal-to-noise accumulation.  
It captures many of the relevant features, including 
the \'{e}tendue, throughput efficiencies, exposure duty cycle, sky brightness, 
and pixel sampling.  

Here we propose a new figure of merit for time-domain surveys, the
instantaneous volumetric survey speed, that is motivated specifically by
transient discovery.  In Section \ref{sec:speed}, we
define the figure of merit and discuss the issue of spectroscopic
accessibility.  In Section \ref{sec:cadence}, we discuss how the related
metric of areal survey rate determines the range of cadences achievable.  
In Section \ref{sec:detection}, we extend this methodology to
compute transient detection rates.

We use \textit{cadence} throughout
the paper to mean generically 
the actual time sequence of exposures obtained by
a survey, including weather losses and daylight for ground-based surveys.  
Survey cadences can thus be irregular or contain
multiple timescales\footnote{For example, the baseline LSST Wide-Fast-Deep
	Survey includes a pair of visits separated by $\sim$30 minutes,
	with the next revisit three nights later.}.  
Our analysis in this paper will focus on strictly \textit{regular}
cadences, in which each field in the survey area is repeatedly 
revisited at cadence intervals $\Delta t$.  Thus a ``one hour cadence''
indicates that each field is visited throughout the night with separations
of one hour, and then returned to on subsequent nights for further visits 
on that same temporal grid.

\section{Speed} \label{sec:speed}

We define the instantaneous 
volumetric survey speed $\dot{V}_M$ as the comoving spatial 
volume in
which an object of fiducial absolute magnitude $M$ may be detected in a
single exposure with specified signal-to-noise ratio (SNR), 
divided by the total time per exposure
(exposure time plus any readout and slew overheads):
\begin{equation}
\dot{V}_M = \frac{\Omega_{\rm fov}}{4\pi}
\frac{V_{\rm c}(z_{\rm lim}(M, t_{\rm exp}))}{t_{\rm exp} + t_{\rm OH}}.
\end{equation}  
In this equation, $\Omega_{\rm fov}$ is the camera field of view, 
$t_{\rm exp}$ and $t_{\rm OH}$ are the exposure and overhead times, and
$V_{\rm c}(z_{\rm lim})$ is the comoving volume as a function of the
redshift of an object at the detection limit $z_{\rm lim}$.  In turn,
$z_{\rm lim}$ depends on the fiducial absolute magnitude $M$ and the limiting
magnitude $m_{\rm lim}$ (and thus $t_{\rm exp}$).  We use
the k-correction of a source with constant spectral density per unit
wavelength $f_\lambda$, $K = -2.5
\log_{10}(1/(1+z))$ \citep{Hogg:99:DistanceMeasures}.  (Using an analytic
k-correction simplifies the computations and enables generic comparisons. 
For true rate estimation, $K$-corrections for specific source
classes should be used when possible.)

This metric
implicitly incorporates many key parameters: the volume depends on 
the field of view of the camera and its limiting magnitude. The limiting
magnitude in turn
depends on the telescope aperture and image quality, filter bandpasses and
throughputs, the local sky background, electronics read noise, pipeline
efficiency, etc.\footnote{Obtaining a limiting magnitude representative of the
true distribution of observing conditions, particularly lunar phase and
seeing,  is vital for useful comparisons between surveys.}  
The time per exposure depends on the configuration and 
performance of the readout electronics and telescope systems.

While we have cast this figure of merit in terms of detection of explosive
transients, it is also relevant for studies of photometrically variable
objects.  If cosmological corrections are small because the volume probed
is local (due to small $M$ and/or $m_{\rm lim}$), maximizing $\dot{V}_M$ also
maximizes the SNR times the number of background-limited sources observed
per unit time.  

We can compare our figure of merit to the capability metric specified
by \citetalias{Tonry:11:ATLAS}.  That metric is composed of
fixed values including the
camera field of view, telescope collecting area, telescope throughput, PSF,
sky background, and duty cycle.  It then relates these to a
trade space of possible survey parameters,
including the SNR at a given
magnitude, the cadence interval, and the total sky area covered per cadence
interval.  However, we can compare the \citetalias{Tonry:11:ATLAS} 
survey metric to our instantaneous
survey speed by evaluating the variable right hand side
of Equation 9 of \citetalias{Tonry:11:ATLAS} for a single exposure:
\begin{equation}
\begin{split}
	{\rm FOM} & = \frac{{\rm SNR}^2\,\Omega_{\rm fov}}{t_{\rm exp} + t_{\rm OH}}
 10^{0.8 m} \\
 & \propto \frac{\Omega_{\rm fov}}{t_{\rm exp} + t_{\rm OH}} (10^{0.2
m_{\rm lim}})^4 \\
& \propto \frac{\Omega_{\rm fov}}{t_{\rm exp} + t_{\rm OH}} d^4 
\end{split}
\end{equation}
for a Euclidean volume where $d = 10^{0.2(m-M+5)}$\,pc.  In contrast, for
non-cosmological events, $V_{\rm c} \propto d^3$, and thus
\begin{equation}
\dot{V}_M \propto \frac{\Omega_{\rm fov}}{t_{\rm exp} + t_{\rm OH}} d^3. 
\end{equation}
So our figure of merit for transient detection scales as
the third power of distance probed, where the \citetalias{Tonry:11:ATLAS}
capability metric derived from SNR accumulation scales as the fourth
power of distance.

\begin{figure}[!htb]
\begin{center}
\includegraphics[width=0.65\textwidth]{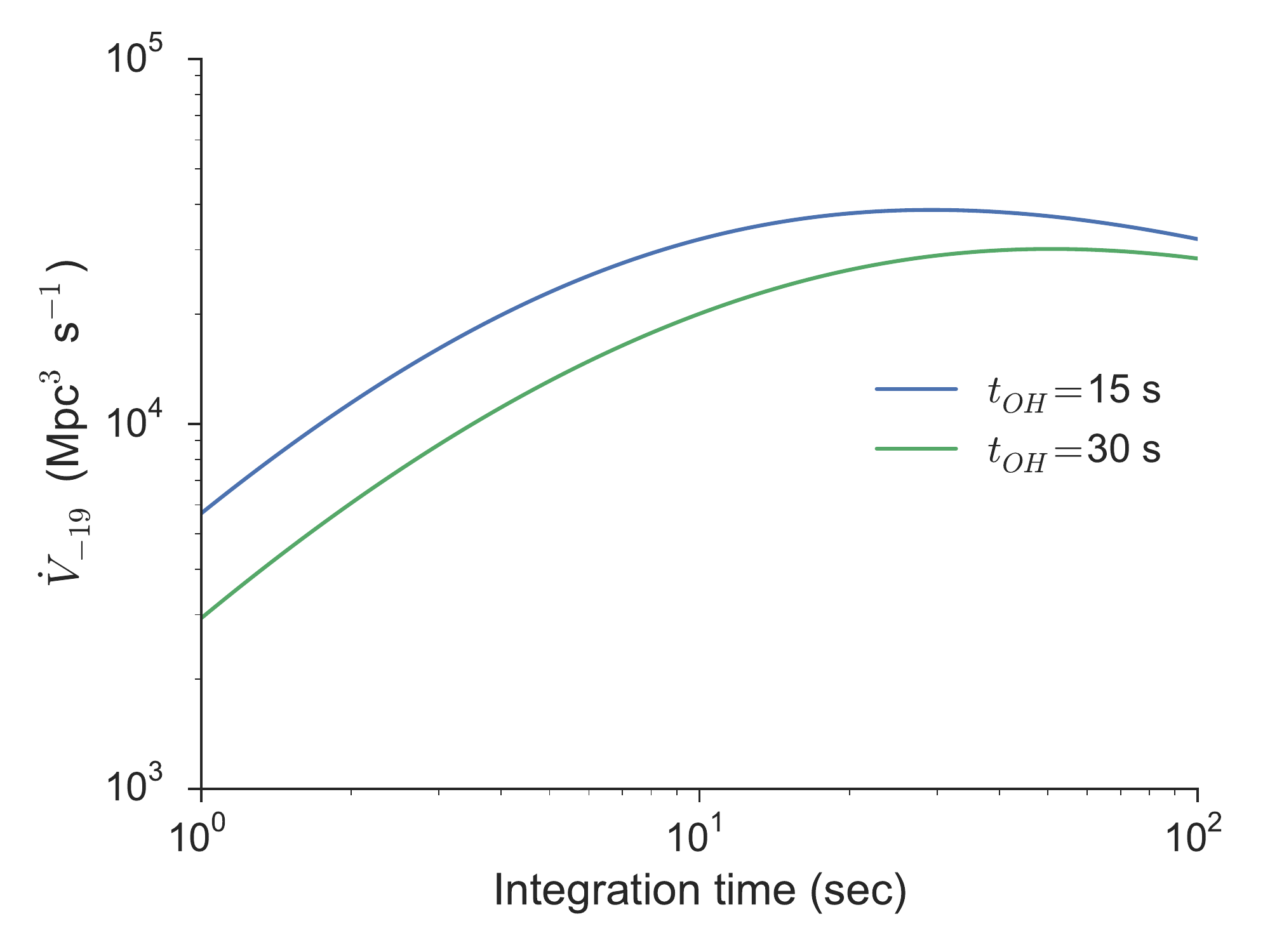}
\caption{Dependence of $\dot{V}_{-19}$ on $t_{\rm exp}$ for ZTF
\citep{Bellm:14:ZTFHotwired} for fiducial
$t_{\rm OH}$ of 15 and 30 seconds.  A longer overhead both decreases the
maximum survey speed achievable and lengthens the optimum exposure time.}
\label{fig:vdot_vs_texp}
\end{center}
\end{figure}

Interestingly, selecting $\dot{V}_M$ as the figure of merit implies that
any specific camera has an optimal exposure time for discovering transient
events.
That optimum depends most strongly on the overhead time
between exposures.  Intuitively, exposure times that are short compared to
the overhead are inefficient. Exposures that are too long increase the
surveyed volume only through an increased single exposure depth ($V \propto
t_{\rm exp}^\frac{1}{4}$), which is less effective than increasing the areal coverage
of the snapshot ($V \propto t_{\rm exp}$).  Given the presence of cosmological
integrals, it is most convenient to find the optimum exposure time
maximizing $\dot{V}_M$  using numerical methods.  Figure
\ref{fig:vdot_vs_texp} shows the dependence of $\dot{V}_M$ on $t_{\rm exp}$
and $t_{\rm OH}$ for a specific camera realization.

For definiteness, we use a fiducial value of $M=-19$ (characteristic of
Type Ia supernovae) throughout this work.   This choice creates some
dependence on cosmology and assumed $k$-correction on the derived value of
$\dot{V}_M$ for the deepest surveys.  We use a cosmology with $h=0.704$,
$\Omega_m = 0.272$, $\Omega_\Lambda = 0.728$ \citep{Komatsu:11:WMAP7BAO} as
implemented in the package
\texttt{cosmolopy}\footnote{\url{http://roban.github.io/CosmoloPy/}}.

The total
spatial volume surveyed in a cadence interval (a ``snapshot'') 
is proportional to the
number of transients in the snapshot.\footnote{
cf.\ Figures 8.5 and 8.10 of the LSST Science Book
\citep{LSST:09:ScienceBook}.  
Strict proportionality requires
that the transients be uniformly distributed throughout the volume
surveyed, which may not be the case for Galactic or local universe
transients, and that confusion does not limit the depth of the exposures as
integration time increases \citepalias[see also][]{Tonry:11:ATLAS}.}
%However, two or more observations of the field may be required during the
%lifetime of the transient in order to discover it (Section
%\ref{sec:detection}).
The figure of merit $\dot{V}_M$ thus describes the
capability of a given observing system to trade the volume surveyed against
revisit time.
%It defines the slope of the snapshot volume vs. cadence interval trade line
%(Figure \ref{fig:snapshot_volume_vs_cadence}).  (Very short and very long cadences require different
%treatment, as discussed in Section \ref{sec:cadence}.)
Maximizing $\dot{V}_M$ when designing a camera thus maximizes its 
ability to discover transients at any desired revisit time, subject to the
constraints on cadence intervals that we will discuss in Section \ref{sec:cadence}.

Not all transients are created equal, however.  Full scientific
exploitation of a detected transient typically requires additional
photometric and spectroscopic followup.  The feasibility of this followup
depends strongly on the apparent magnitude of the transient.  A survey
discovering a smaller absolute number of transients may thus be more
productive if those transients are brighter and can be observed with
more readily available moderate-aperture followup telescopes.

Accordingly, we define a modified figure of merit, the
{\em spectroscopically-accessible} volumetric survey speed:

\begin{equation}
\dot{V}_{M, m < s} = f_{\rm spec}(s) \dot{V}_M,
\end{equation}
where $f_{\rm spec}(s)$ is the fraction of the comoving volume producing
transients with apparent magnitudes brighter\footnote{This is equivalent
to using the brighter of the survey's limiting magnitude and $s$ when
computing $z_{\rm lim}$.  It assumes that the volume where $m > s$ is not
useful for transient detection.  This is an oversimplification, as faint
early detections can provide valuable information for nearby transients later
peaking at brighter apparent magnitudes.  However, depending on the cadence,
coaddition of several shallow exposures may fill this role.} than $s$.
We choose $s$ based on the capability of the followup resources available:
$s \approx 21$ is a reasonable limit for observations with 
3--5\,m telescopes, while $s \approx 23$ is reasonable for 8--10\,m followup.

This scheme of defining $f_{\rm spec}$ with a sharp cutoff at apparent
magnitude $s$ assumes our priority is to be capable of following up the
faintest (presumably rare) transients.  If instead we wish to obtain a
large sample of transient spectra, it will be more useful to weight the
comoving volume integral by the cost (in time) of followup as a 
function of apparent
magnitude\footnote{For single-object spectroscopy with fixed target
acquisition time
$t_{\rm ac}$ and fiducial exposure time $t_0$ for objects of apparent
magnitude $m_0$, this weighting is $t_0 10^{0.8 (m- m_0)} + t_{\rm ac}$
divided by the length of the night.}.  
This weighting will further emphasize the strengths of the
wide, shallow surveys producing the most bright transients.

Because the sharp cutoff at apparent magnitude $s$ is conceptually simpler, 
we use it through the remainder of this work.  We use $5\sigma$ limiting
magnitudes throughout.

\begin{figure}[!htb]
\begin{center}
\includegraphics[width=\columnwidth]{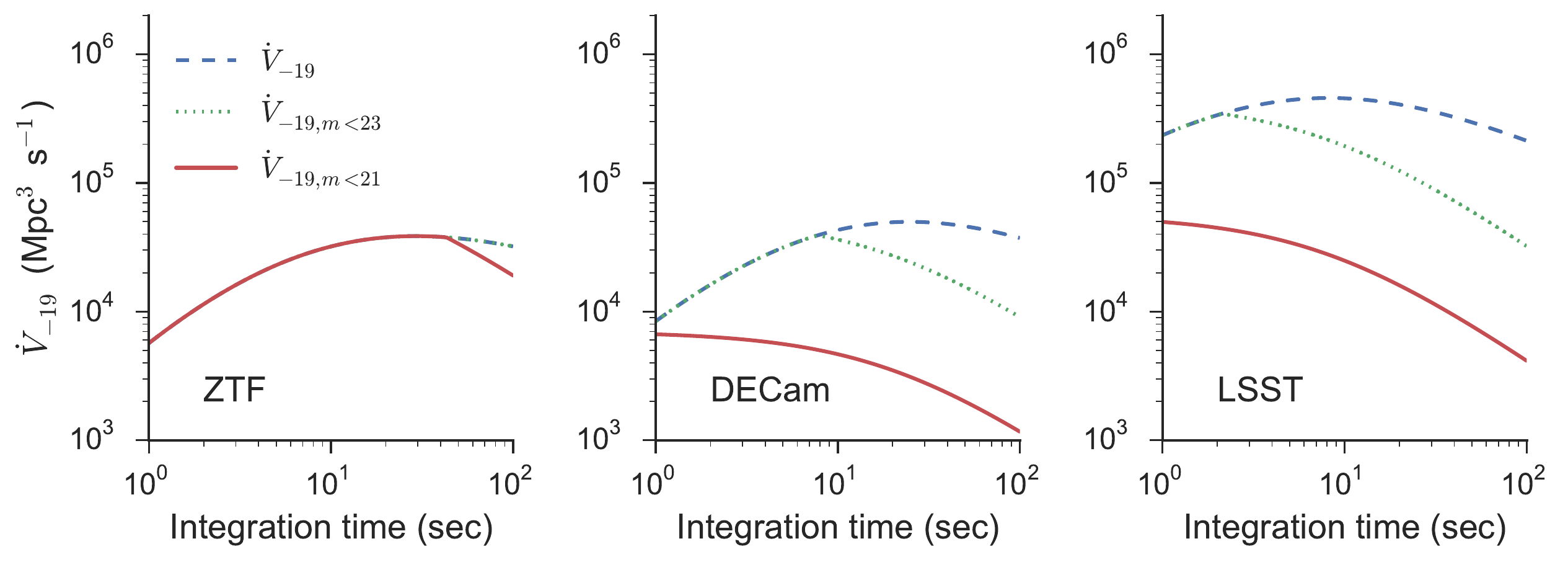}
\caption{Variation in spectroscopically-accessible survey speed with 
exposure time for ZTF, DECam, and LSST.   Solid red lines show  
the volume probed for transients brighter than 21$^{\rm st}$\,mag 
($\dot{V}_{-19, m < 21}$), dotted green lines show 
$\dot{V}_{-19, m < 23}$, and dashed blue lines show the total volumetric
survey speed ($\dot{V}_{-19}$).  Larger aperture telescopes may be less
efficient at detecting bright transients even at short exposure times.
}
\label{fig:spec_vdot_by_telescope}
\end{center}
\end{figure}

\begin{figure}[!htb]
\begin{center}
\includegraphics[width=\textwidth]{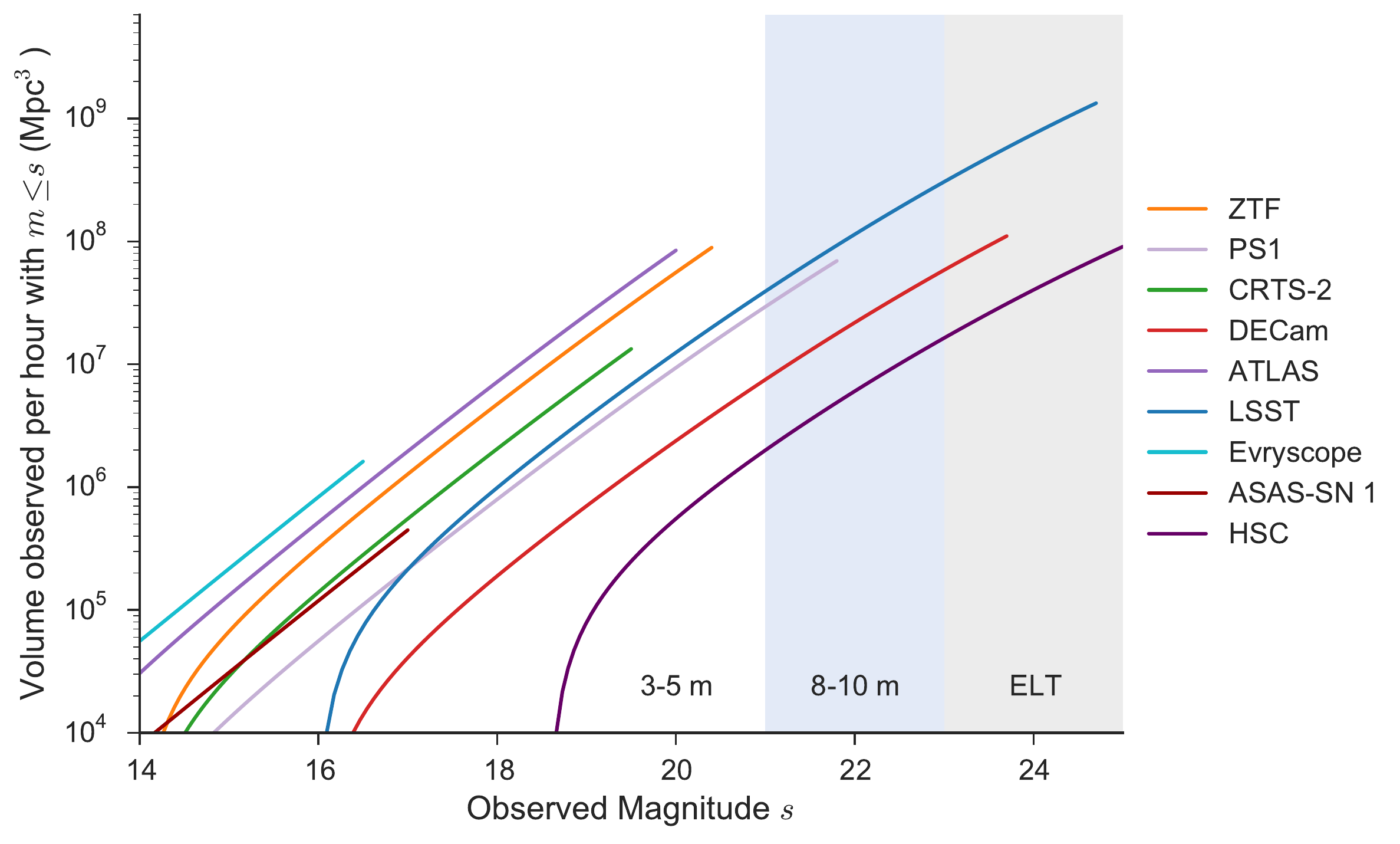}
\caption{Spatial volume within which transients of absolute magnitude $-19$
may be detected in a one-hour survey snapshot versus the maximum 
apparent magnitude $s$ of the detection.
Shaded regions indicate the telescope aperture necessary for spectroscopic
followup.}
\label{fig:vdot_vs_obs_mag}
\end{center}
\end{figure}

Table \ref{tab:surveys} lists instrument specifications and the resulting
survey speeds for several major time-domain surveys.
Figure \ref{fig:spec_vdot_by_telescope} shows the impact of the limiting magnitude cut on $\dot{V}_M$
and on the optimal exposure time.
Figure \ref{fig:vdot_vs_obs_mag} shows the 
spatial volume surveyed as a function of transient brightness.

% vdot vs texp w/ s cuts.  Have to code in all the telescope pars--maybe
% ZTF and decam?

%%%%%%%%%%%%%%%%%%%%%%%%%%%%%%%%%%%%%%%%%%%%%%
% use Surveys.py/wrap_survey_speeds to get numbers
\setlength{\tabcolsep}{4pt}
\begin{table}
{\small 
\centering
\begin{tabular}{|l|ccccccc|cc|cc|}
\hline
Survey&
$D$ &
$\Omega_{\rm fov}$&
Etendue &
Pixels &
$t_{\rm exp}$ &
$t_{\rm OH}$ &
$m_{\rm lim}$ &
$\dot{\Omega}$ &
$N_{\rm obs}$ & 
$\dot{V}_{-19}$ &
$f_{\rm spec}$ \\  
Camera &
{\small (m)} &
{\small (deg$^2$)} &
{\small (m$^2$ deg$^2$)} &
{\small ($^{\prime\prime}$)} &
{\small (sec) }  &
{\small (sec)}  &
&
{\small (deg$^2$\,hr$^{-1}$)} &
{\small (yr$^{-1}$)} &
{\small (Mpc$^3$/s)} &
 \\
\hline
Evryscope  & 0.06(27$\times$) & 8660 & 26.5 & 13.3 & 120 & 4 & 16.4 & 251419 & 19279 & \num{ 1.1e+04 } & 1.00 \\
ASAS-SN 1 & 0.14(4$\times$) & 73 & 1.1 & 7.8 & 180 & 23 & 17 & 1294 & 99 & \num{ 1.2e+02 } & 1.00 \\
ATLAS & 0.5(2$\times$) & 60 & 11.8 & 1.9 & 30 & 8 & 20.0 & 5684 & 435 & \num{ 2.3e+04 } & 1.00 \\
CRTS & 0.7 & 8.0 & 3.1 & 2.5 & 30 & 18 & 19.5 & 600 & 46 & \num{ 1.4e+03 } & 1.00 \\
CRTS-2 & 0.7 & 19.0 & 7.3 & 1.5 & 30 & 12 & 19.5 & 1628 & 124 & \num{ 3.7e+03 } & 1.00 \\
LSQ & 1.0 & 8.7 & 6.8 & 0.9 & 60 & 40 & 20.5 & 313 & 24 & \num{ 2.3e+03 } & 1.00 \\
% 8.65 deg^2
PTF & 1.2 & 7.3 & 8.2 & 1.0 & 60 & 46 & 20.7 & 246 & 18 & \num{ 2.3e+03 } & 1.00 \\
Skymapper & 1.3 & 5.7 & 7.5 & 0.5 & 110  & 20  & 21.6 & 157 & 12 & \num{ 3.9e+03 } & 0.52 \\
% 1.13 equivalent unobstructed aperture; 5.68 deg^2
PS1 3$\pi$ & 1.8 & 7.0 & 17.8 & 0.3 & 30  & 10 & 21.8 & 630 & 48 & \num{ 1.9e+04 } & 0.42 \\
% MDS uses 250 sec exposures; 3Pi uses 30 sec
SST & 2.9 & 6.0 & 39.6 & 0.9 & 1 & 6 & 20.7 & 3085 & 236 & \num{ 2.7e+04 } & 1.00 \\
MegaCam & 3.6 & 1.0  & 10.2 & 0.2 & 300 & 40  & 22.8 & 10 & 0.8 & \num{8.8e+02} & 0.16 \\
DECam & 4.0 & 3.0 & 37.7 & 0.3 & 50 & 20  & 23.7 & 154 & 11 & \num{ 2.9e+04
} & 0.07 \\
HSC & 8.2 & 1.7 & 89.8 & 0.2 & 60 & 20 & 24.6 & 76 & 5 & \num{ 3.1e+04 } & 0.03 \\
%& see http://www.naoj.org/Projects/HSC/index.html and http://www.naoj.org/cgi-bin/img\_etc.cgi \\
\hline
BlackGEM$^*$ & 0.6(4$\times$) & 2(4$\times$) & 11.3 & 0.6 & 30 & 5 & 20.7 & 822 & 63 & \num{ 7.6e+03 } & 1.00 \\
ZTF$^*$ & 1.2 & 47 & 53.1 & 1.0 & 30 & 15  & 20.4 & 3760 & 288 & \num{ 2.5e+04 } & 1.00 \\
LSST$^*$ & 6.7 & 9.6 & 319.5 & 0.2 & 30 & 11 & 24.7 & 842 & 64 & \num{ 3.7e+05 } & 0.03 \\
% 8.4m real D, 6.67m effective filled aperture
\hline

\end{tabular}
\caption{{\small Comparison of existing and planned wide-field
optical survey cameras.  For each camera we list telescope aperture
($D$), single-image field of view ($\Omega_{\rm fov}$), etendue, pixel
scale,
integration time ($t_{\rm exp}$), overhead per exposure ($t_{\rm
OH}$), the $5\sigma$ single-exposure limiting magnitude in $r$ ($m_{\rm lim}$),
the areal survey rate ($\dot{\Omega}$),
the number of observations
per field per year in a hypothetical uniform $3\pi$ survey averaging
6.5 hours per night observing time ($N_{\rm obs}$),
the volume probed per unit exposure time for transients of absolute
magnitude $-19$ ($\dot{V}_{-19}$), and the fraction of $\dot{V}_{-19}$ that
is ``spectroscopically accessible'' ($m \le 21$\,mag; $f_{\rm spec} \equiv
\dot{V}_{-19,m<21}/\dot{V}_{-19}$)
Performance for future ($^*$) surveys is estimated.
BlackGEM values are for Phase 1 (4 telescopes); ASAS-SN values are for a
single site. CRTS values are for the CSS telescope only.
SST and LSST diameters are effective apertures.
References: 
Evryscope: \citet{Law:15:EvryscopeScience};
ASAS-SN: \citet{Shappee:14:ASASSN}, B.\ Shappee, priv.\ comm.;
ATLAS: \citet{Tonry:11:ATLAS,Tonry:13:PS1ATLAS}, J.\ Tonry priv.\ comm.; 
CRTS: \cite{Drake:09:CatalinaFirstResults}; 
CRTS-2: A.\ Mahabal, priv.\ comm.;
LSQ (La Silla QUEST): \citet{Rabinowitz:12:LSQ};
PTF: \citet{Law:09:PTFOverview};
Skymapper: \citet{Keller:07:SkyMapperOverview};
PS1: \citet{Kaiser:04:PanSTARRs}, \citet{Morganson:12:PS1Quasar};
SST: \citet{Freedman-Woods:14:SST}, \citet{Ruprecht:14:SST};
MegaCam: \citet{Boulade:03:MegaCam},
\url{http://www.cfht.hawaii.edu/Instruments/Imaging/MegaPrime/quickinformation.html};
DECam: \citet{DePoy:08:DECam}, NOAO Data Handbook;
HSC: \citet{Miyazaki:12:HSC}, \citet{Tanaka:16:HSCTransients}, HSC E.T.C.;
BlackGEM: \url{http://astro.ru.nl/blackgem/};
ZTF: \citet{Bellm:14:ZTFHotwired};
LSST: \citet{LSST:09:ScienceBook}.
}
\label{tab:surveys}}
}
\end{table}
\setlength{\tabcolsep}{6pt}

\section{Cadence} \label{sec:cadence}

In Section \ref{sec:speed}, we showed the effectiveness of wide, shallow
surveys in detecting spectroscopically-accessible transients.  However, 
the proportionality between $\dot{V}_M$ and the number of detected
transients breaks down if a survey runs out of new sky to observe.  For
modern wide-field surveys, it is easily feasible to observe the entire
visible sky in less than one night.  We therefore must consider the relationship
between a survey's {\em areal} survey speed $\dot{\Omega} =
\frac{\Omega_{\rm fov}}{t_{\rm exp} + t_{\rm OH}}$, its latitude $\phi$, and the
possible cadences. 

In this section, we consider an idealized and simplified transient survey.
We assume that our survey operates in a single filter bandpass at a single
site\footnote{We treat surveys using multiple telescopes at one or more sites
closely spaced geographically (e.g., ATLAS, Evryscope, PanSTARRS 1 \& 2) 
as single
instruments with the combined fields of view of all telescopes.  We here
consider only single sites of widely-separated surveys (e.g., ASAS-SN North
and South) because of the additional complexity of treating the field
overlap regions.}
with no weather losses.  While observing, we observe the largest
snapshot area ($\Omega_{\rm snap}$) possible in the cadence
interval ($\Delta t$) such that we can observe the entire footprint a
second time in the second epoch.  We assume a single exposure time
(optimized for the cadence interval chosen if necessary) and a fixed overhead
between exposures, implying roughly constant slews between each 
exposure\footnote{We assume generically that each field is observed only
once per cadence interval, but paired exposures without slews may be
accommodated in this scheme by summing the resulting exposure and overhead
times.}.  Finally, we limit our observations in the footprint to times when
the fields are above a specified maximum airmass or zenith angle ($\zeta_{\rm
max}$).

The trade space between survey snapshot area $\Omega_{\rm snap}$ and
cadence interval 
$\Delta t$ has two limits.  The first is when an instrument sits on
a single field and takes exposures at a rate 
limited only by its readout time.  In
this case $\Omega_{\rm snap, min} = \Omega_{\rm fov}$ and $\Delta t_{\rm
min} = t_{\rm OH}$.  
Surveys operating at this limit are usually driven by specialized
science goals; they are best undertaken by instruments with extremely large
fields of view (such as Pi of the Sky \citep{Burd:05:PioftheSky} or Evryscope
\citep{Law:14:Evryscope})
and/or fast readout time (EM-CCDs or CMOS).

The opposite limit is to maximize the snapshot volume, and hence use the
longest cadence interval possible.
The maximum revisit time ($\Delta t_{\rm max}$) 
is set by how long it takes a survey with areal survey speed $\dot{\Omega}$
to cover the entire visible sky area.
The limit of the ``available sky'' thus
depends on the observatory latitude, which determines the length of the
night as well as the rotation rate of new sky into the observable region
above $\zeta_{\rm max}$.

\begin{figure}[!htb]
\begin{center}
\includegraphics[width=0.65\textwidth]{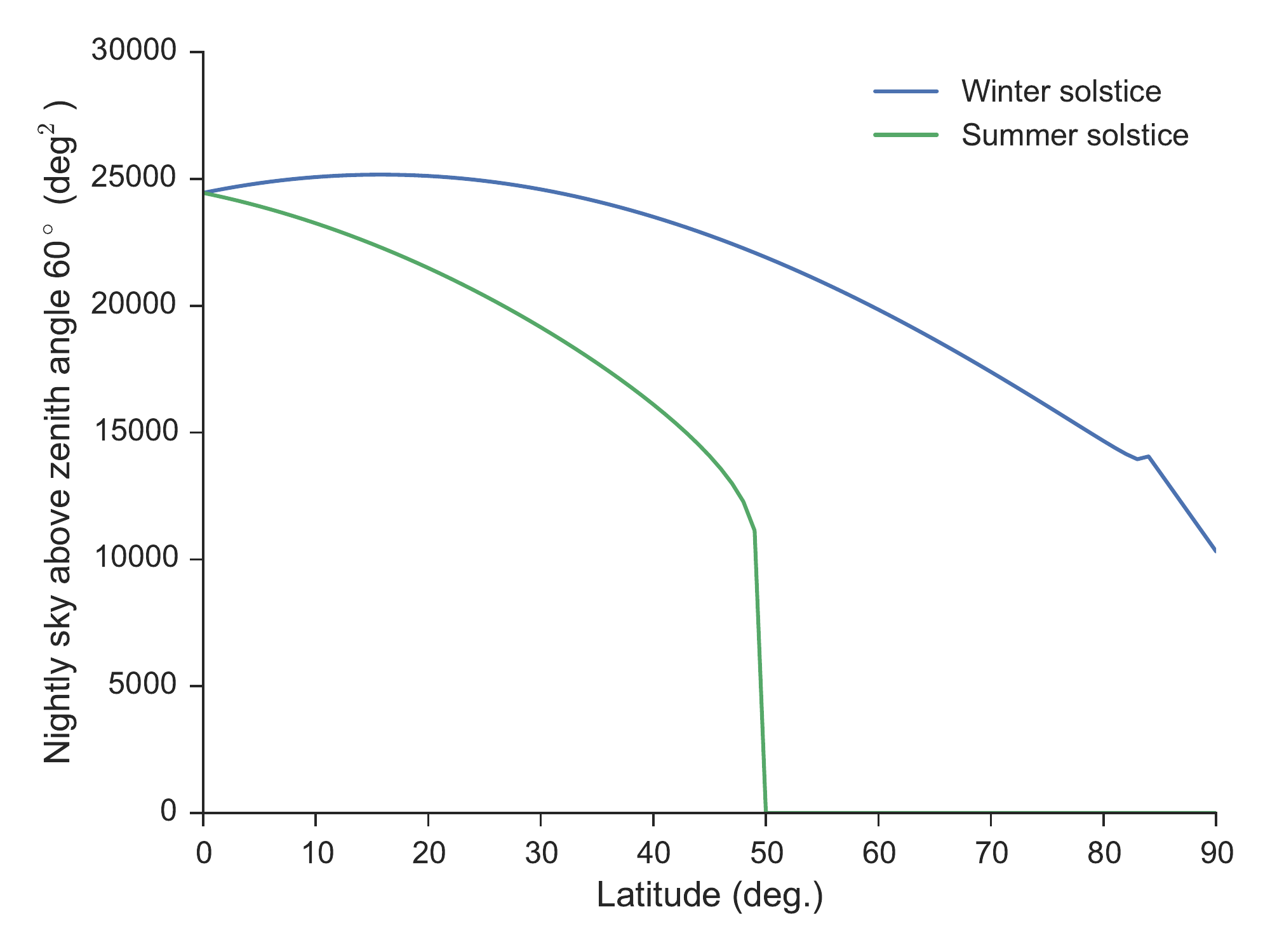}
\caption{Unique sky area per night above $\zeta_{\rm max} = 60^\circ$ as a 
function of observatory latitude.  Night lengths are determined for 18$^\circ$
twilight, and the limiting values at the summer and winter solstices are
shown.  The cusp near 84$^\circ$ latitude is due to the transition to
24\,hrs of $>18^\circ$ darkness.
\label{fig:unique_sky_per_night}
}
\end{center}
\end{figure}

Calculating the limiting cadence interval $\Delta t_{\rm max}$ requires
consideration of several cases.  The sky area above the zenith angle cut 
$\zeta_{\rm max}$ at any given instant may be divided into a circumpolar region
and a region that will rotate below $\zeta_{\rm max}$ eventually:
\[
\Omega_{\rm inst} = \Omega_{\rm circ} + \Omega_{\rm r}
= 2\pi (1 - \cos \zeta_{\rm max}).
\]
(Depending
on the latitude and $\zeta_{\rm max}$, there may be no circumpolar region or
the entire sky may be circumpolar.)  Since the circumpolar region stays
above the zenith angle cut, the rate of change of this instantaneous sky is 
\[
\frac{d \Omega_{\rm inst}}{dt} = \frac{d \Omega_{\rm r}}{dt} 
\]
The rotation of sky into and out of $\Omega_{\rm r}$ is most easily
calculated by integrating the areal rotation across the
meridian\footnote{Given the sidereal rotation rate $\dot{H}$, $\frac{d
\Omega_{\rm r}}{dt} = \dot{H} (\cos \theta_1 - \cos \theta_2)$, where the
limits of the integration $\theta_1, \theta_2$ are set by the colatitude
$\varphi = 90 - \phi$ and $\zeta_{\rm max}$.  If $\varphi >= \zeta_{\rm max}$,
there is no circumpolar area, and the limits of integration are
$\theta_{1,2} = \varphi \pm \zeta_{\rm max}$.  Otherwise $\theta_1 = \zeta_{\rm
max} - \varphi$ and $\theta_2 = \varphi + \zeta_{\rm max}$: the length of
the meridian above $\zeta_{\rm max}$ but outside the circumpolar region.
(To simplify the presentation, we restrict to Northern latitudes.)
}.
The total sky area passing above $\zeta_{\rm max}$ in one night is therefore
\[
\Omega_{\rm night} = \Omega_{\rm inst} + \frac{d \Omega_{\rm r}}{dt} \Delta
t_{\rm night}.
\]
Figure \ref{fig:unique_sky_per_night} shows the dependence of the total sky
available per night on observatory latitude.

The first case of limiting cadences to consider is for instruments capable
of surveying the entire sky in less than a single night.  These must have
areal survey rates $\dot{\Omega} > \frac{d \Omega_{\rm r}}{dt}$: they must
be able to survey the sky in the region above $\zeta_{\rm max}$ faster than it
rotates out of the available field.  In this case $\Omega_{\rm snap} =
\Omega_{\rm inst}$: we choose a footprint such that at
the end of the first epoch,
the trailing edge has risen to $\zeta_{\rm max}$ to be observed and 
the leading edge of the footprint has just
rotated down to $\zeta_{\rm max}$ to be observed in the second epoch.
The limiting cadence interval is thus $\Delta t_{\rm max} = \Omega_{\rm
inst}/\dot{\Omega}$.  The remaining check is to ensure that 
$\Delta t_{\rm max}$ is less than half of the night length.

In cases where $\dot{\Omega} < \frac{d \Omega_{\rm r}}{dt}$ or it takes
longer than half a night to survey $\Omega_{\rm inst}$, it will take more
than one night to repeat observations of the available sky above
$\zeta_{\rm max}$.
For instruments with $\dot{\Omega}$ greater than 
$\frac{d \Omega_{\rm r}}{dt}$ scaled to the nightly sidereal rotation,
the argument is identical to the
sub-night case. We replace $\Omega_{\rm inst}$ by $\Omega_{\rm night}$
averaged over the cadence interval
and restrict the observing time within $\Delta t$ to the times the sun is
down.  Given the dependence on night length, $\Omega_{\rm snap,max}$ and
$\Delta t_{\rm max}$ are most conveniently found numerically.

Instruments with $\dot{\Omega}$ slower than the sidereal rotation of
the footprint are unlikely to be used for time domain surveys, so we do not
consider them further.

\begin{figure}[!htb]
\begin{center}
\includegraphics[width=0.75\textwidth]{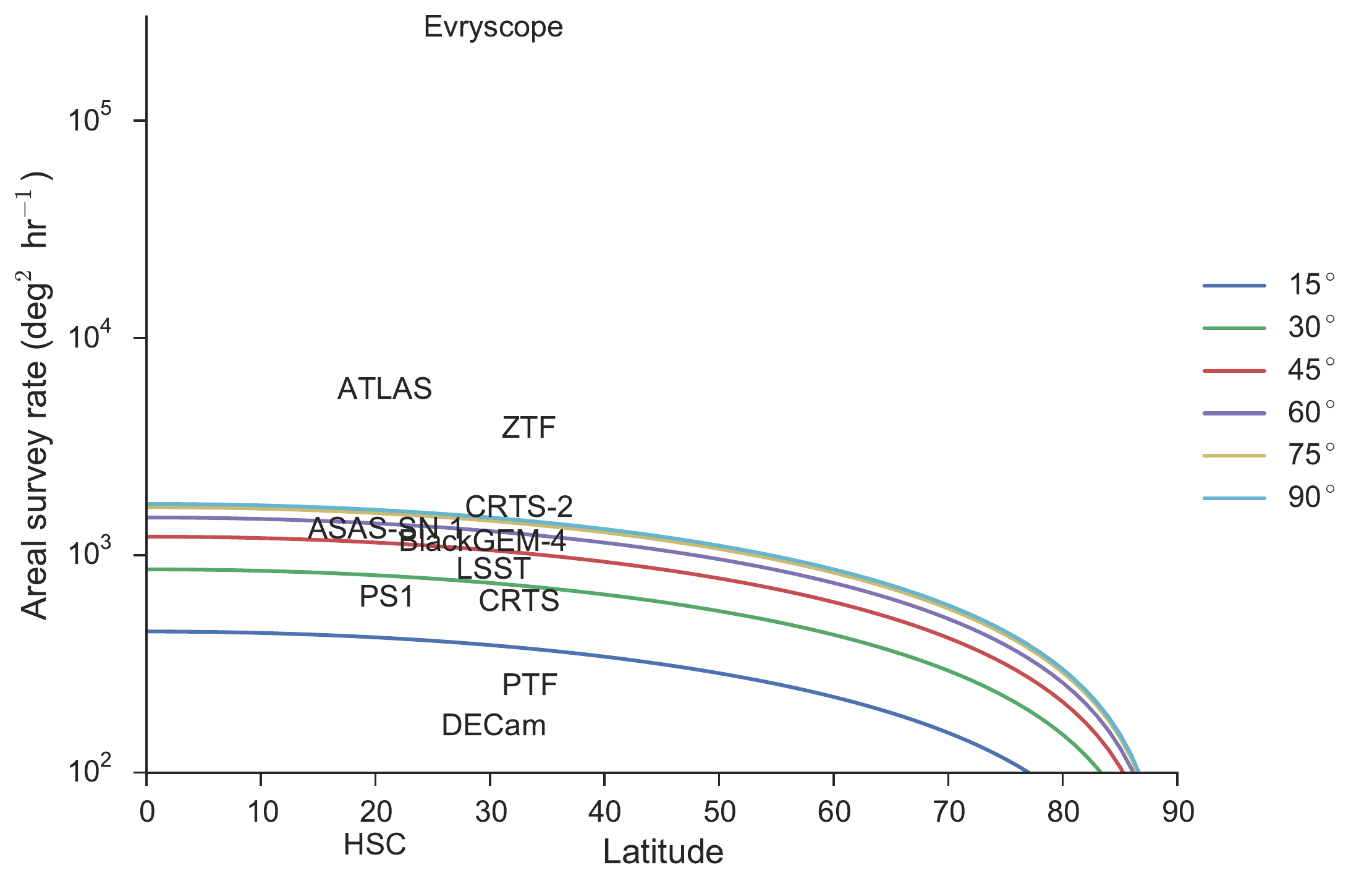}
\caption{Areal survey rates ($\dot{\Omega}$, in deg$^2$ hr$^{-1}$) for
specific surveys.
Overplot colored lines are the rate
of change of sky area above a given zenith angle cut $\zeta_{\rm max}$ as a
function of latitude ($\frac{d \Omega_{\rm r}}{dt}$, in deg$^2$ hr$^{-1}$).  
Surveys above a given line can survey faster than the sky rotates within
the footprint defined by $\zeta_{\rm max}$.
%The dotted line shows the nightly change in accessible sky area above 
%$z_{\rm max} = 90^\circ$ due to sidereal rotation.
\label{fig:areal_survey_rate_vs_rot}
}
\end{center}
\end{figure}

Figure \ref{fig:areal_survey_rate_vs_rot} plots the areal survey speeds of several cameras against the 
footprint rotation rate $\frac{d \Omega_{\rm r}}{dt}$.

%todo: areal survey rate against max snapshot area/cadence time?

\begin{figure}[!htbp]
\begin{center}
\includegraphics[width=1.0\textwidth]{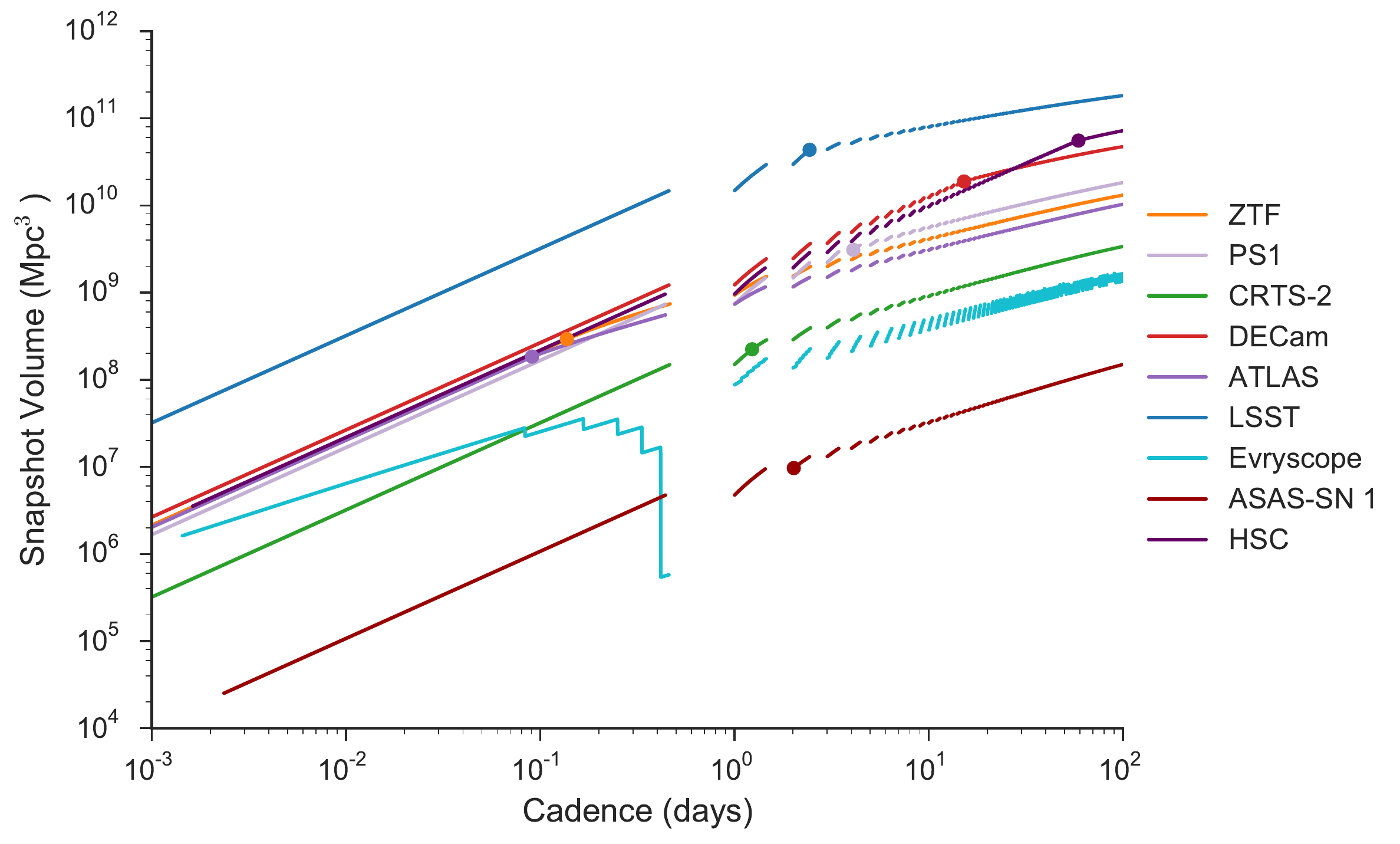}
\caption{Total snapshot survey volume for $M=-19$ transients 
versus cadence interval for several surveys.  
Observations begin on the winter solstice.  
Filled circles indicate the cadence interval 
at which the survey runs out of available sky
above an airmass of 2.5.  Below this point the snapshot volume is
simply the survey speed $\dot{V}_M$ times the available observing time.
Above this point we increase the exposure time to reach longer cadence
intervals, at cost of a slower rate of increase in the snapshot volume.
\label{fig:snapshot_volume_vs_cadence}
}
\end{center}
\end{figure}

\begin{figure}[!htbp]
\begin{center}
\includegraphics[width=1.0\textwidth]{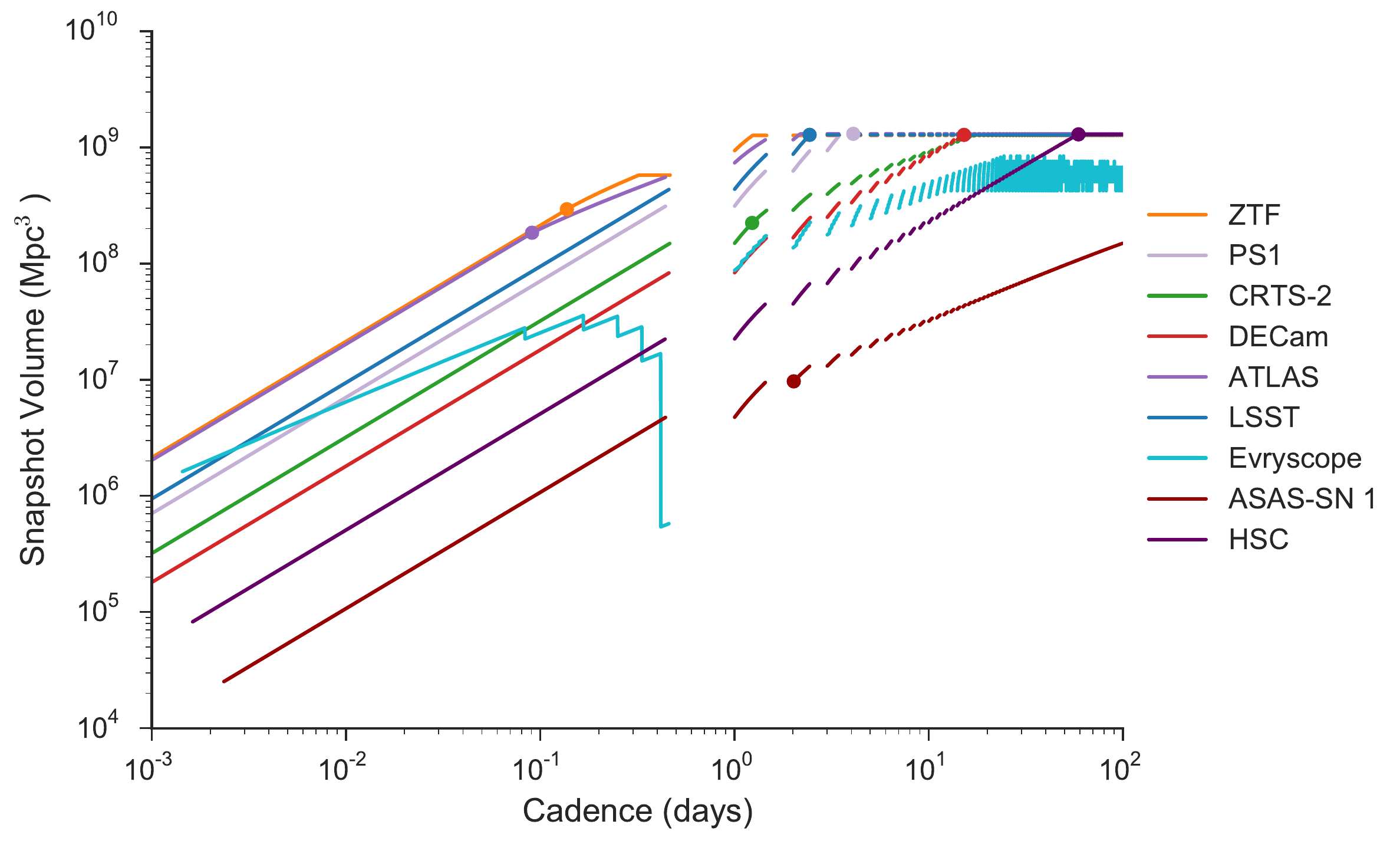}
\caption{Same as Figure \ref{fig:snapshot_volume_vs_cadence}, but for
spectroscopically-accessible volume only ($\dot{V}_{M, m < 21}$): we consider only the volume in
which we can detect transients of apparent magnitude brighter than $m =
21$, appropriate for 3--5\,m-class spectroscopic followup.  ZTF and ATLAS
will discover more bright transients at any chosen cadence interval than even
LSST.
\label{fig:accessible_volume_vs_cadence}
}
\end{center}
\end{figure}

Figures \ref{fig:snapshot_volume_vs_cadence} and
\ref{fig:accessible_volume_vs_cadence} plot the snapshot volume against the
cadence interval, with and without a cutoff on the transient limiting magnitude.

For some cameras, the longest sky-limited cadence interval $\Delta t_{\rm max}$ 
may be shorter than the transient timescale of interest.  For example,
superluminous supernovae (SLSN) may be visible for hundreds of days; 
two all-sky surveys with identical depths using a one day and a one week
cadence would thus each discover SLSN at the same rate.
Several modifications of our
baseline survey are possible in this case.  A first option is to maintain
the higher cadence, achieving finer sampling of the lightcurve and hence
improved characterization of the lightcurve shape.  This may
be scientifically valuable in many cases, although it does not increase the
transient detection rate.  A second option is to integrate for longer
exposures than needed to maximize $\dot{V}_M$, as in the sky-limited case
the ``optimal'' exposure time no longer maximizes the number of detected
transients.  However, with deeper exposures the additional 
transients discovered will be more
challenging to follow up.  Survey extensions are also possible: surveys
in other filters or of other programs can productively fill time 
before beginning a new epoch.

\section{Detection} \label{sec:detection}

The selection of a specific cadence interval $\Delta t$ sets the volume 
$V_{\rm snap}(M)$ within which transients of absolute magnitude $M$ may be detected.  
It also imposes a selection effect on the decay timescale of the transients
detected.  In particular, events which decay much more quickly than $\Delta
t$ are unlikely to be detected.

Exact computation of detection rates requires detailed modeling of
multi-color lightcurves, detection passbands, cosmological evolution of
event rates, event-to-event variations, and more
\citep[e.g.,][]{Kessler:09:SNANA}.
Our goal in this work is to provide a reasonable comparison of survey
camera capability and broad cadence tradeoffs rather than 
a precise estimate of
the rates of specific event types.  
Accordingly, we make several simplifying assumptions to enable analytic
integrals.  
However, it is straightforward to
extend this methodology to specific event classes by substituting
appropriate lightcurve shapes, k-corrections, evolution of the
rates with redshift, and extinction.

We calculate the yearly detection rate for transients of absolute magnitude
$M$ and rest-frame effective decay timescale $\tau_{\rm eff}$ 
by integrating over the co-moving volume:
% TODO: parse a little more carefully the proportionality (or not) of V_M
% to detection rate?

\begin{equation}
N(M,\tau_{\rm eff}, m_{\rm lim}, \Delta t) = \Omega_{\rm snap} \int_0^{z_{\rm lim}(M, m_{\rm lim})} \!
\frac{\mathcal{R}(z)}{1+z} ct_k(M, m_{\rm lim}, \tau_{\rm eff}, z)
\frac{dV}{dz} dz.
\label{eqn:ndets}
\end{equation}
Here $\mathcal{R}(z)$ is the comoving volumetric rate (events Mpc$^{-3}$
yr$^{-1}$); we divide by (1+$z$) due to time dilation.  
The survey limiting magnitude determines the depth of the
spatial cone probed, while the choice of cadence interval 
$\Delta t$ determines its
angular extent $\Omega_{\rm snap}$.

We calculate the effect of observing cadence on the discovery rate using
the control time $ct_k$.  Here $k$ is the
number of consecutive images in which we require a detection.
If a given transient
is detectable in the observer frame above $m_{\rm lim}$ 
for time $t_{\rm vis}$, given an array of
separations $dt_i$ between $n$ consecutive images of a field, we define $r$ as
the number of intervals where $dt_i < t_{\rm vis}$.  Then the control time
is defined by \citet{Zwicky:42:ControlTime} as 

\begin{equation}
ct_1(t_{\rm vis}) = \sum_1^r dt_i + (n - r) t_{\rm vis}.
\end{equation}
Using the control time in this manner counts even a single
detection of an event above the limiting magnitude towards the total number
of events discovered ($k=1$).  
In modern transient surveys it is common to require
multiple detections of an event before triggering followup in order to
avoid contamination by uncatalogued asteroids and image subtraction
artifacts.  We develop here an extension of the $ct_1$ algorithm for $k >
1$ (Algorithm \ref{alg:ctk}).

\begin{algorithm}
\caption{Algorithm for computing the control time $ct_k$ from $n$
observations with timestamps $t_i$, requiring $k > 1$ consecutive observations
within the visibility interval $t_{\rm vis}$.}
\label{alg:ctk}
\begin{algorithmic}
\Begin
\For{j:=0}{n-k} 
\Begin
\State $dt_j = t_{j+k-1} - t_j$ 
\If {dt_j > t_{\rm vis}}
\State $ct_j = 0$
\Else
\If {ct_{j-1} = 0 {~\rm \bf or~} j = 0}
\State $ct_j = t_{\rm vis}$
\Else
\State $ct_j = dt_j$
\End
\State $ct_k = \sum_{j=0}^{n-k} ct_j$
\End.
\end{algorithmic}
\end{algorithm}

%In practice this means that Equation \ref{eqn:ndets}
%is an overestimate, as most transient surveys require multiple consecutive 
%detections to confirm a discovery and trigger followup 
%in order to filter out uncataloged asteroids.

We use a simple analytic approximation for the lightcurve shape to simplify 
the calculation of the control time.   We assume that the transient rises
and falls linearly in magnitude space with characteristic rest-frame timescales
$\tau_{\rm rise}$ and $\tau_{\rm fall}$ days mag$^{-1}$.  
Accordingly, the source is visible in the observer frame for 

\begin{align}
t_{\rm vis}(M, m_{\rm lim}, \tau_{\rm rise}, \tau_{\rm fall}) & = 
   ( (M - m_{\rm lim}) \tau_{\rm rise} + (M - m_{\rm lim}) \tau_{\rm
   fall} ) (1+z) \\
   & \equiv (M - m_{\rm lim}) \tau_{\rm eff} (1 + z).
\end{align}

For many explosive transients, $\tau_{\rm fall} >> \tau_{\rm rise}$, so 
$\tau_{\rm eff} \approx \tau_{\rm fall}$.

Using these assumptions, we may now compute (idealized) detection
rates for different surveys and cadences.  We calculate the detection rates 
in a grid of transient peak magnitude $M$ and effective timescale
$\tau_{\rm eff}$ \citep[cf.][]{Kasliwal:11:Thesis}.  We use a constant fiducial
volumetric rate density $\mathcal{R}(z) = 3\times10^{-5}$ events 
Mpc$^{-3}$ yr$^{-1}$, approximately the local SN Ia rate.  As above, we
use the k-correction of a hypothetical $f_\lambda$ standard.

For each cadence interval $\Delta t$, we create a grid of times for a one-year
interval with spacing $\Delta t$.  We mask all grid points between
eighteen degree dawn and eighteen degree twilight and use the remainder as
our observation times $t_i$ to compute the control time\footnote{ 
In reality, various points in the survey footprint
will be surveyed between $t_i$ and $t_i + \Delta t$.  This gridded approach
is simple to compute and will be a good approximation if the order in which
the fields are observed is consistent from night to night.  
Assessing detection rates for complex pointing schemes and 
including weather losses
requires the use of a full survey simulator, which is beyond the scope of
this work.}.  As discussed in Section \ref{sec:cadence}, when the desired
cadence interval is longer than the time needed to survey the available sky 
above an airmass of 2.5
we increase the exposure time to compensate (cf. Figure
\ref{fig:snapshot_volume_vs_cadence}).
% we are interested in contributions of purely geometric effects & survey
% camera/telescope parameters

\begin{figure}[!htbp]
\centering
\includegraphics[width=0.75\textwidth]{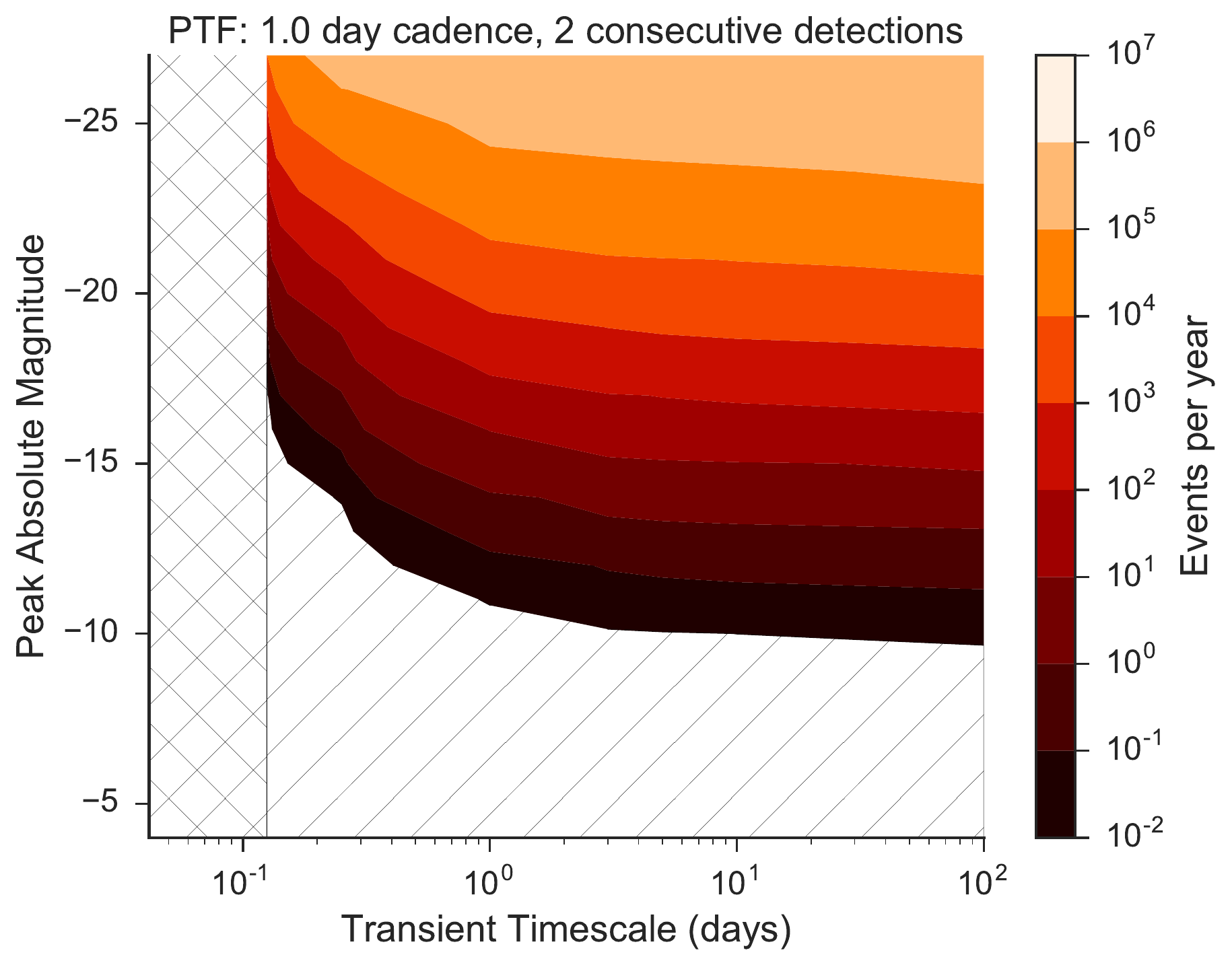}
\caption{
Number of transients detectable in at least two epochs by
PTF using a strict 1 day cadence with no weather losses as a function of
peak absolute magnitude and effective timescale $\tau_{\rm eff}$.
Colors indicate the number of events detected per year assuming all
transients occur at the local SN Ia rate.
Transients in the hashed region are detected at a rate of less than one per
century, while events in the crosshatched region cannot be detected twice
by the specified cadence.}
\label{fig:ptf1day}
\end{figure}

Figure \ref{fig:ptf1day}
shows the detection rate for PTF using a 1 day cadence in the phase space
of transient peak magnitude and effective timescale.
Comparing the predicted numbers of detections at timescales less than a day
to the paucity of fast transients discovered to date
\citep[cf.][]{LSST:09:ScienceBook, Kasliwal:11:Thesis} 
emphasizes that any fast transients that
exist must be rare, as current surveys already have some sensitivity to
short-timescale events.
We also could easily invert the calculation to determine the volumetric rate
$\mathcal{R}(z)$ compatible with current nondetections.

\begin{figure}[!htbp]
\centering
\includegraphics[width=0.75\textwidth]{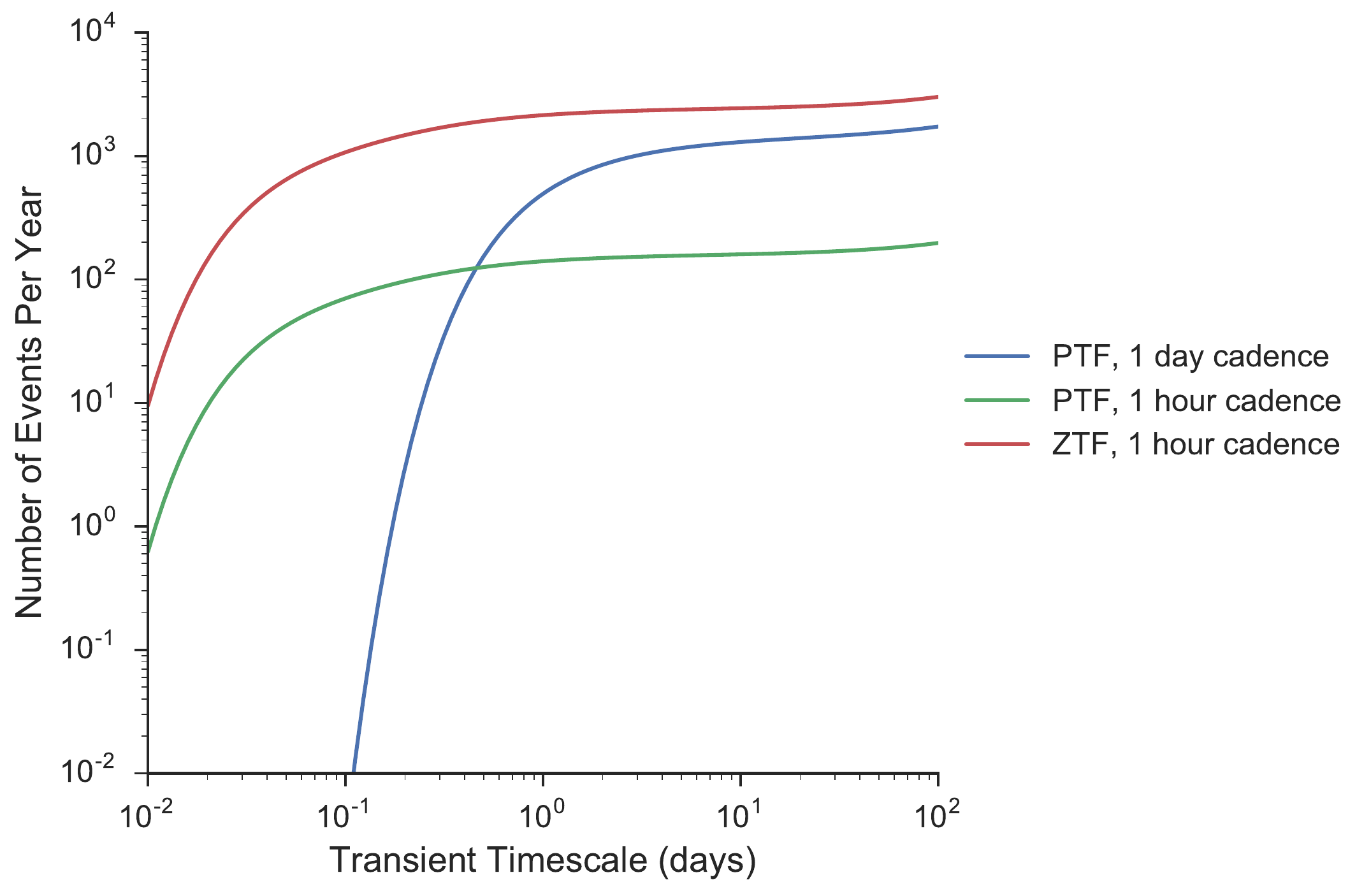}
\caption{
Comparison of the number of $M=-19$ transients detectable 
by PTF and ZTF as a function of cadence interval 
and effective timescale $\tau_{\rm eff}$.
The requirement to detect the transient 
in at least two epochs 
leads to a sharp drop in the number of events detected below the cadence
interval $\Delta t$.
PTF can increase its sensitivity to (intrinsically rare)
fast transients by observing at a 1
hour cadence rather 1 day cadence, but this decreases the number of slow
transients it detects by an order of magnitude.  Thanks to its much 
greater survey speed $\dot{V}_{-19}$, ZTF can discover more transients than
PTF at all timescales $\tau_{\rm eff}$ even with a 1 hour cadence.  }
\label{fig:ptfvsztf}
\end{figure}

In Figure \ref{fig:ptfvsztf}, 
we slice Figure \ref{fig:ptf1day} at the fiducial value
of $M = -19$ mag and compare PTF 1 day and 1 hour cadences to ZTF at a 1
hour cadence.
By increasing the raw survey
speed relative to PTF, ZTF can break from the cadence--survey volume trade 
space defined by PTF and conduct a survey that is both wide area and high
cadence.  Such a survey is required to discover intrinsically 
rare, fast-decaying events such as gamma-ray burst afterglows.

%For a
%source class of interest (e.g., SN Ia),  
%the cadence interval and limiting
%magnitude determine the fraction of transients within this volume which may
%be detected.  This detectable fraction $f_{\rm det}$ may be estimated by
%sampling from an ensemble of light curves within $\Delta t$ of the peak
%brightness (or from the explosion date, if young transients are of
%interest).  
% use detectable fraction, or sum over volumes using the range of absolute
% magnitudes enountered?  would need f(M'): fraction of transients with M > M' 
% within delta t of peak, then do an f-weighted sum of the volume.
%The number of detected transients per year is then the volumetric transient
%rate (in Mpc$^3$ yr$^{-1}$) times $V_c f_{\rm det}$ times the 
%yearly exposure.

%examples: SNe Ia, LGRB afterglows.
% list number of transient class detected at identical cadences by
% different instruments
% plot detection rate of a transient class vs cadence interval?

\begin{figure}[!htbp]
\centering
\includegraphics[width=0.75\textwidth]{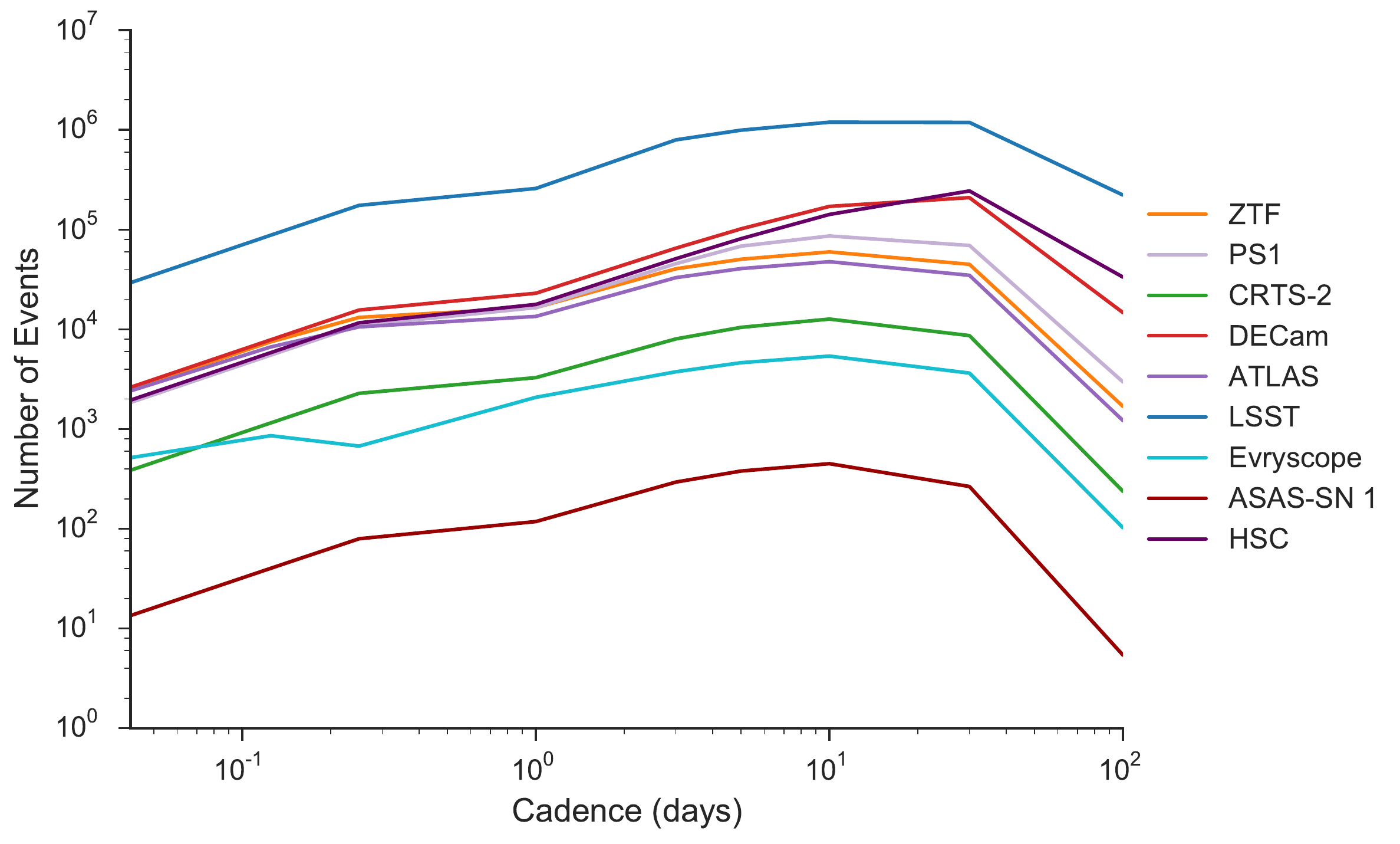}
\caption{Number of $M=-19, \tau_{\rm eff}=20$\,days\,mag$^{-1}$ transients
	detected in at least two epochs by various surveys 
	as a function of cadence interval, 
	assuming a constant volumetric rate equal
	to the local SN Ia rate.  
At short cadences ($\Delta t << \tau_{\rm eff}$), 
high $\dot{\Omega}$ surveys like ZTF and ATLAS can discover comparable
numbers of transients to deeper surveys like DECam and HSC.
At long cadences  ($\Delta t >> \tau_{\rm eff}$), the detection efficiency
of all surveys declines because all but the brightest nearby events decay
too quickly to be observed in a second epoch.
\label{fig:ndets_bycadence_20}}
\end{figure}

\begin{figure}[!htbp]
\centering
\includegraphics[width=0.75\textwidth]{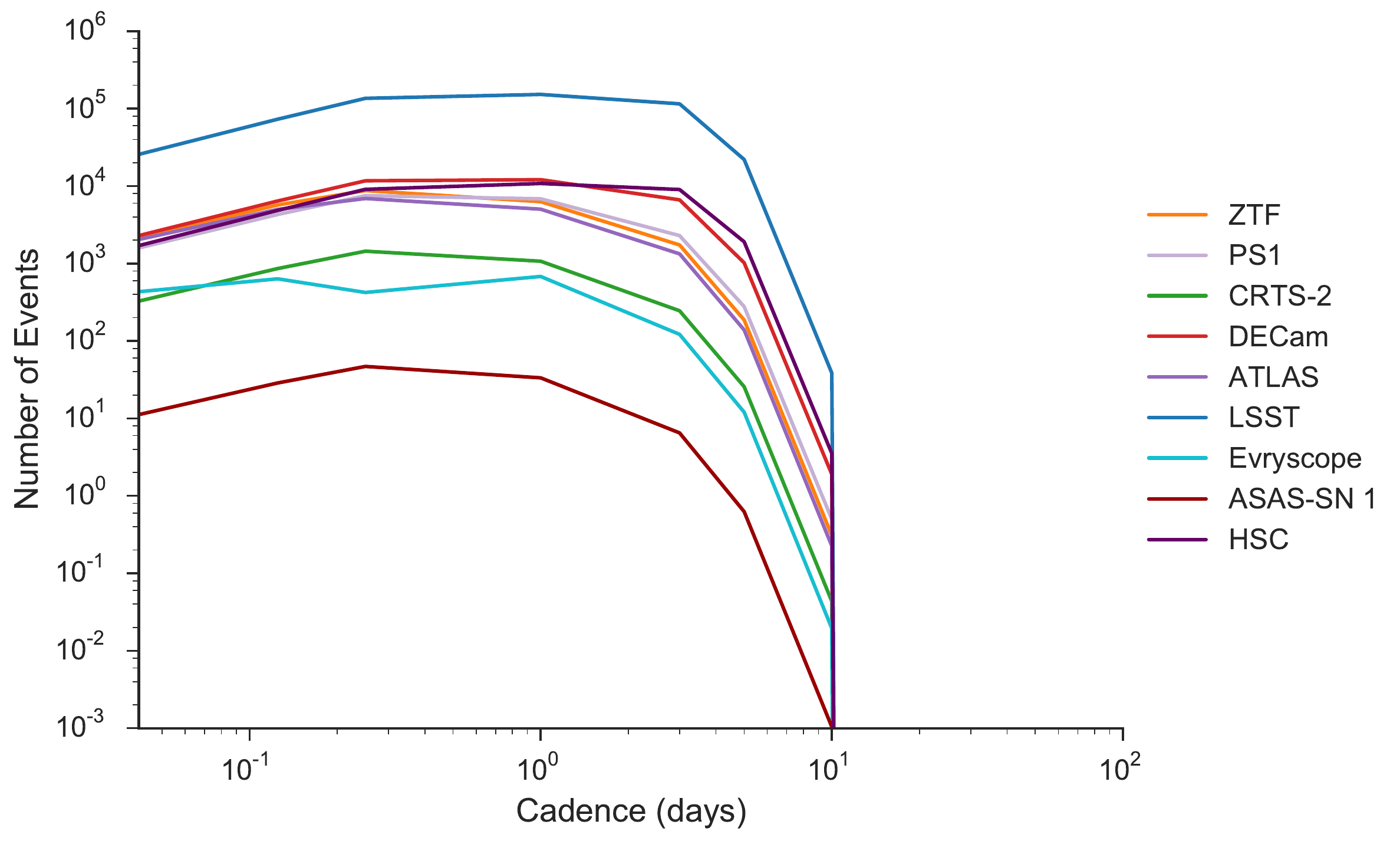}
\caption{Number of $M=-19, \tau_{\rm eff}=1$\,day\,mag$^{-1}$ transients
	detected in at least two epochs by various surveys.
\label{fig:ndets_bycadence_1}}
\end{figure}

With the ability to estimate the transient detection rate as a function of
cadence, it is then possible to choose a cadence interval to optimize the total
number of detections for one or several source classes.
Figures \ref{fig:ndets_bycadence_20} and \ref{fig:ndets_bycadence_1} show the dependence of the detection rate on
the chosen survey cadence for decay rates $\tau_{\rm eff} = 20$ and 1
days\,mag$^{-1}$.  


\section{Conclusion}

To obtain useful comparisons of transient surveys, it is necessary to look
beyond simple calculations of \'{e}tendue.
We have developed a new means of comparing current and near-term time-domain
surveys: the instantaneous volumetric survey speed.
This metric requires only readily-available information: the camera field of view, exposure and
overhead times, and limiting magnitude.  It captures the relationship
between the cadence interval and the survey snapshot volume, which is related
to the discovery rate.  The volumetric survey speed is straightforward to
interpret physically, and additionally it implies an optimal integration
time.

A closely-related metric is the areal survey rate, which serves to bound
the achievable cadences for a ground-based survey.  Simply put, many modern
time-domain surveys run out of fresh sky to survey, sometimes in a
single night!  
Another practical concern is the brightness of the
discovered transients---for science applications requiring spectroscopic
followup, discovering many faint transients is often less valuable than a 
finding a few bright ones.

We have developed a basic analytic framework for estimating the
detection rate of transients that evolve at different speeds.  By
assessing the influence of the survey parameters and the chosen survey
cadence on the detection rate,
one may optimize the survey cadence for the science of interest and compare
to other surveys.  
A complete evaluation of detection rates for specific transient types will
require analysis of actual simulated or realized pointing histories, with
increased fidelity coming at the cost of additional complexity.
LSST's Metrics Analysis Framework \citep{Jones:14:MAF} 
provides one such means of performing
quantitative assessments of pointing histories.

Maximizing the transient detection rate does not alone optimize a survey
design.  In many cases, there is tension between the need for
well-sampled transient lightcurves and the desire to maximize the 
discovery rate.
Practical limits such as the availability of followup resources,
theoretical or systematic limitations, or multiple scientific objectives
may also apply.  However, quantitative assessment of these tradeoffs and of the
competitive landscape will strengthen the design of transient surveys. 

\acknowledgments

The author thanks Shri Kulkarni, Tom Prince, Eran Ofek, Paul Groot, and the
anonymous referee for conversations and suggestions that improved this work.

\bibliographystyle{yahapj}
\bibliography{ptf}

\end{document}